# Electronic mean free path in as-produced and purified single-wall carbon nanotubes

H. Kajiura,[a] A. Nandyala, U. C. Coskun, and A. Bezryadin

*Department of Physics, University of Illinois at Urbana-Champaign,*

*Urbana, Illinois 61801*

M. Shiraishi,[b] and M. Ata [c]

*Materials Laboratories, Sony Corporation, Atsugi-city, Kanagawa 243-0021, Japan*

[a] To whom correspondence should be addressed.

Fax: +81-45-226-3773. Tel: +81-45-46-2453.

Electronic addresses: hisashi.kajiura@jp.sony.com

Postal address: Materials Laboratories, Sony Corporation, 4-16-1 Okata, Atsugi-city, Kanagawa, 243-0021, Japan

b) Present address, Graduate School of Engineering Science, Osaka University.

c) Present address, National Institute of Advanced Industrial Science and Technology (AIST).

Abstract

The effect of purification on room temperature electronic transport properties of single-wall carbon nanotubes (SWNT) was studied by submerging samples into liquid mercury. The conductance plots of purified SWNTs showed plateaus, indicating weak dependence of the electrical resistance on the length of the tube connecting the electrodes, providing evidence of quasi-ballistic conduction in SWNTs. The electronic mean free path of the purified SWNTs reached a few microns, which is longer than that of the as-produced SWNTs, and which is consistent with the calculation based on the scattering by acoustic phonons. (Applied Physics Letters, vol. 86, 122106 (2005).)

Measurements of electronic transport in single-wall carbon nanotubes (SWNT)[1,2] have shown a potential for various applications: *e.g.* nano-sized electronic devices.[3,4,5,6,7,8,9] To obtain optimal performance, it is desirable to use SWNTs with as a low level of disorder as possible. A common method of SWNT production is laser ablation,[3] but as-produced nanotubes are covered with impurities such as amorphous carbon, which act as scattering centers in electronic transport. We found that a purification process[10] leads to an improvement of SWNTs' transport properties. In this study, we report how this process affects transport properties, especially the electronic mean free path (EMFP), of SWNTs. To determine the EMFP of individual nanotubes, we submerged samples into liquid mercury (Hg) and measured the variation in conductance.[11] We find that the EMFP is much longer in the purified samples than in as-produced ones. The purified samples show quasi-ballistic electronic transport with the EMFP reaching a few microns, which is consistent with the calculation based on the scattering by acoustic phonons.[9]

The SWNT soot was synthesized using a Ni/Co catalyst by laser ablation at 1200$^{o}$C.[3,10] Transmission electron microscopy (TEM, HF2000, Hitachi) showed that SWNTs in the as-produced soot were covered with impurities, *e.g.* amorphous carbon (Fig. 1a). The as-produced soot was purified using $H_2O_2$, HCl and NaOH solutions and heated at 650$^{o}$C at a pressure of $10^{-2}$ Pa for 1 h.[10] The purified soot consisted of clean-surface SWNTs with a diameter of 1.4 nm (Fig. 1b). Scanning electron microscopy (SEM, S4700, Hitachi) revealed SWNT-ropes protruding from the soot (Fig. 1c). Lovall *et al.* reported that a single nanotube protrudes from the tip of a SWNT-rope.[12]

Our measurements, which provide the transport properties of a single SWNT, were made using a piezo-driven nanopositioning system at room temperature in air (Fig. 2a).[13] A piezo-positioner allowed gentle and reproducible contact between the nanotubes and the Hg counter electrode.[11] The soot sample was attached to a metallic mobile electrode (probe) using silver paste. The probe was then attached to the piezo-positioner with a displacement range of 20 μm (17PAZ005, MELLES GRIOT). To make electrical contact between the sample and Hg, the probe was driven cyclically up and down with a peak-to-peak amplitude of 2-10 μm and a frequency of 0.1-1 Hz. A potential of 180 mV was applied between the probe and Hg electrodes. The current was measured as a function of piezo-positioner displacement with a typical sampling rate of 1000 points/s using an analog-to-digital converter (NI6120, National Instruments).

Figure 3a shows a conductance plot $G(x)$ of the purified SWNT sample, normalized by the conductance quantum unit, $G_0 \equiv 2e^2/h = (12.9 \text{ k}\Omega)^{-1}$, where $e$ is the elementary charge and $h$ is Planck's constant. Here $x$ represents the piezo-positioner extension, with $x = 0$ corresponding to the point at which the tube-Hg contact is made. Thus $x$ measures the extent the nanotube segment is submerged into Hg (Fig. 2b).[8] In most of the measurements, we observe a sequence of clear steps and plateaus on the $G(x)$ plots. We interrupted each step on the $G(x)$ as a new nanotube made contact with Hg.[11] The appearance of a plateau after each step indicates that the conductance is essentially independent of the length of the SWNT segment connecting the electrodes, implying that the SWNTs behave as quasi-ballistic conductors even at room temperature.

If a *diffusive* metallic wire is connected to two bulk electrodes, the resistance of the system is proportional to its length. On the other hand, with a one-dimensional *ballistic* quantum wire, the resistance of the system is $(nG_0)^{-1}$, independent of the length. Here $n$ is the number of conduction channels ($n = 2$ for metallic carbon nanotubes). With ballistic transport, conductance does not depend on the wire length, and thus our observation of essentially flat plateaus on the $G(x)$ plot provides evidence for quasi-ballistic transport in purified SWNTs at room temperature. Although a $2G_0$ conductance is expected for an ideal metallic nanotube having perfect contact with electrodes,[14] we did not observe the $2G_0$ conductance step. The step size of $1G_0$[11] was also not observed. In our experiment, the first step was at most $0.6G_0$. This deviation from theory can be explained using contact effects. In our setup, the nanotube that touches Hg may not directly contact the mobile probe. Instead, it is connected to other tubes in the soot, so the contact resistance between the probe and the measured tube is high and random.[8] We believe that this contact resistance is responsible for the small size of the first conductance step and causes the step size to vary from sample to sample.

If the tube-Hg contact resistance, $R_{T\text{-}Hg}$, significantly contributes to the total resistance, the conductance plot will have a rounded shape close to $x = 0$.[15] For the as-produced SWNTs, no rounded shape was observed, while the purified SWNTs showed a rounded shape (Fig. 3a). This difference might be due to the difference in surface condition: the purified SWNTs were chemically modified during purification. The estimated $R_{T\text{-}Hg}$ for the purified SWNTs is ~10 Ωμm, which is less than one-tenth of that of multi-wall carbon nanotubes.[15] This low $R_{T\text{-}Hg}$ implies that the contribution of $R_{T\text{-}Hg}$ to the total resistance is negligible. We thus assume that the $R_{T\text{-}Hg}$ is independent of $x$ in both the as-produced and purified samples.

Using the first plateau in the $G(x)$, we estimated the resistance per unit length ($\rho$) of the SWNTs. For this, the $G(x)$ was converted into the resistance $R(x) = 1/G(x)$ (Fig. 3b). The resistance can be approximated as $R(x) = R_c - \rho x$, where $R_c$ is contact resistance.[15] The linear fitting in Fig. 3b provides $R_c$ = 51 kΩ and $\rho$ = 2.4 kΩ/μm. The $R_c$ value obtained from 28 traces measured on four different purified soot samples falls in the range 20-100 kΩ. The $\rho$ of the carbon nanotube is related to the EMFP ($l$) as $\rho = (h/4e^2)(1/l)$.[7] This equation establishes that the EMFP of the purified SWNTs falls in the range 0.4-9.7 μm. Figure 4a shows a histogram of the obtained EMFP, in which the distribution peak is found at ~1.5 μm, which is consistent with the reported value measured using an atomic force microscope (1.6 μm).[9] The as-produced SWNTs (25 traces out of five different soot samples) had a higher $R_c$ (150-1000 kΩ) and a shorter EMFP (0.01-2.2 μm) compared to those of the purified SWNTs. In the EMFP histogram of the as-produced samples (Fig. 4b), the distribution peak is found at ~0.3 μm. Thus the strong enhancement of the electronic mean free path in purified tubes is confirmed.

The differences in conduction properties between the as-produced and the purified samples can be explained in terms of impurity effects. In the as-produced samples, electrons have to pass through an impurity layer at the tube-to-tube junctions, which raises the contact resistance between the probe and the measured nanotube. The scattering caused by the impurities shortens the EMFP. This finding is further confirmed by measurements using commercially available purified SWNTs sample (HiPco-SWNT, Carbon Nanotechnologies, Inc.),[16] which have clean-surface like our purified samples. The HiPco-SWNT samples (25 traces out of three different soot samples) had an $R_c$ (50-150 kΩ) and an EMFP (0.2-13 μm), and the EMFP histogram had a peak at ~1.5 μm (Fig. 4c). These results are consistent with those of our purified SWNT samples.

To determine whether the experimental results are consistent with the theoretical mean free path, we calculated the electron acoustic-phonon scattering rate[9] under the assumption that the tube has a metallic armchair structure with a (10,10) chiral index, corresponding to a diameter of 1.37 nm. The electron–acoustic phonon scattering rate $1/\tau_{ac}$ can be expressed as $\frac{1}{\tau_{ac}} = \frac{8\pi^2}{h} \Xi^2 (\frac{k_B T}{2\sigma v_s^2}) \frac{1}{h v_F} = \frac{v_F}{l_{ac}}$ ,where $\Xi \approx 5 eV$ is the deformation potential (Ref. 9 and eq. (5) in Ref. 17), $k_B$ is Boltzmann's constant, $T$ is the temperature, $\sigma = 3.2 \times 10^{-15}$ kg/m is the nanotube mass per unit length, $v_s = 1.5 \times 10^4$ m/s is the acoustic phonon velocity, $v_F = 8 \times 10^5$ m/s is the Fermi velocity, and $l_{ac}$ is the electron acoustic-phonon scattering mean free path.[9] Using this formula, we obtain $l_{ac} \approx 2$ μm, which is in good agreement with the distribution peaks found in the purified SWNT and the HiPco-SWNT samples (Fig. 4). Thus we conclude that in these samples the EMFP is mainly limited by scattering by acoustic phonons.

In summary, we measured the variations in the conductance of SWNTs by submerging as-produced and purified SWNT samples into Hg. The results show that purification leads to a significant improvement in the EMFP, reaching a few microns at room temperature. The obtained EMFP is consistent with the results of the calculation based on scattering by acoustic phonons.

List of Captions

FIG. 1. TEM images of (a) as-produced and (b) purified SWNTs. (c) The SEM image shows SWNT-ropes protruding from the purified nanotube soot.

FIG. 2. Schematic diagrams of (a) experimental setup and (b) tube-Hg contact. Here $x$ is the nanotube segment submerged into Hg.

FIG. 3. Data obtained from the purified sample. (a) Conductance plot $G(x)$ normalized by $G_0 \equiv 2e^2/h$ and (b) the resistance plot $R(x)$ of the first plateau as a function of piezo-positioner displacement ($x$), measured as the tube is pushed into Hg. The straight line in (b) is the fit given by $R(x) = C - \rho x$. Here $C$ is one fitting parameter ($C \approx R_c$) and $\rho$ is the other fitting parameter representing the resistance per unit length of the tube. In this example $C$ = 51 k$\Omega$ and $\rho$ = 2.4 k$\Omega$/μm.

FIG. 4. Histogram of electronic mean free path for (a) purified SWNTs, (b) as-produced SWNTs, and (c) commercially available purified SWNTs (HiPco-SWNT) supplied by Carbon Nanotechnologies, Inc.

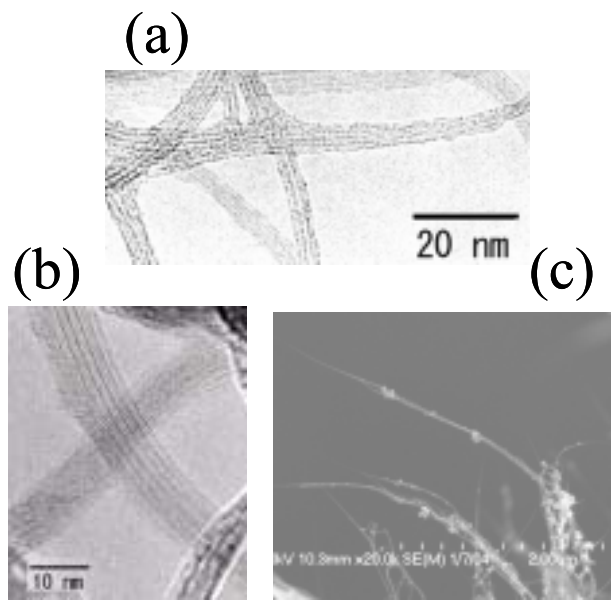

FIG. 1.

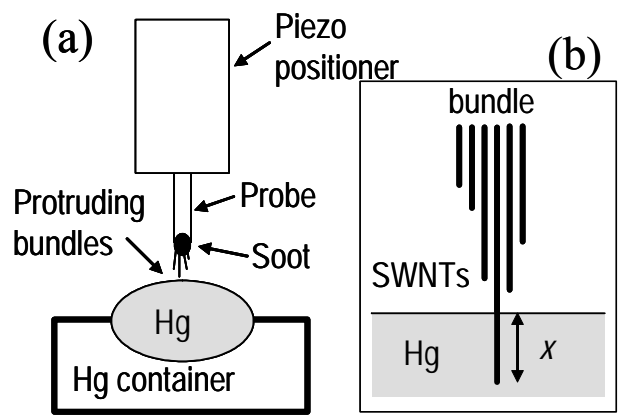

FIG. 2.

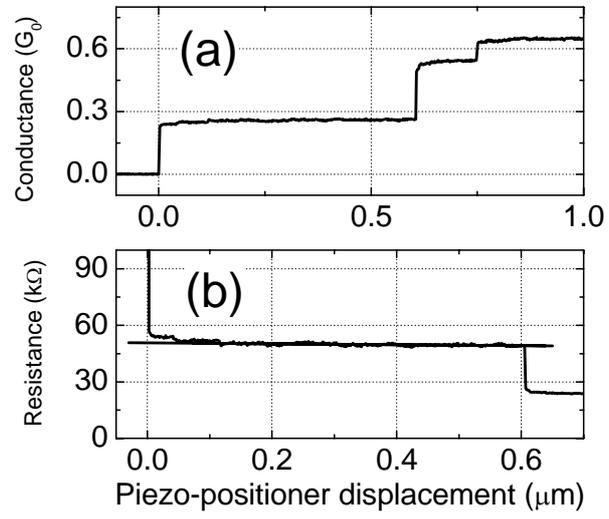

FIG. 3.

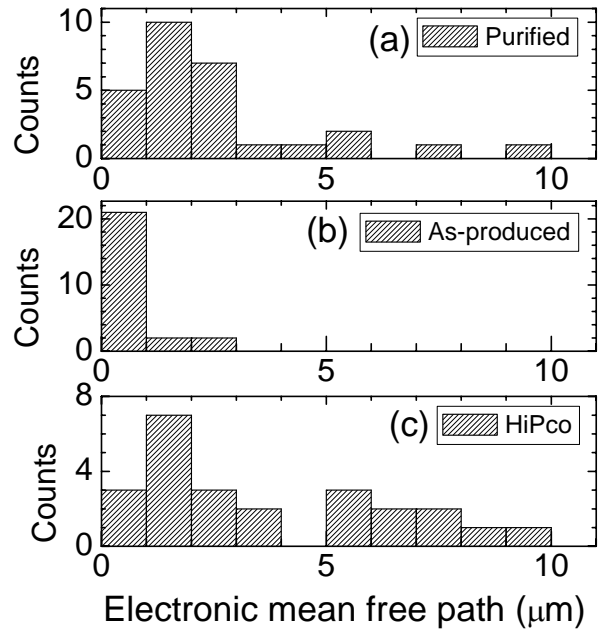

FIG. 4.